\documentclass[%
reprint,
superscriptaddress,
showpacs,
nobibnotes,
 amsmath,amssymb,
 aps,
nopreprintnumbers,
]{revtex4-1}
\pdfoutput=1
\usepackage{graphicx}
\usepackage{dcolumn}
\usepackage{bm}

\usepackage{color}

\begin{document}

\preprint{APS/123-QED}

\title{Neutrino-nucleus quasi-elastic and 2p2h interactions up to 10 GeV}

\author{R. Gran}
\affiliation{Department of Physics, University of Minnesota -- Duluth,
Duluth, Minnesota 55812, USA}
\author{J. Nieves}
\affiliation{Instituto de F\'isica Corpuscular (IFIC), Centro Mixto
  CSIC-Universidad de Valencia, Institutos de Investigaci\'on de
  Paterna, Apartado 22085, E-46071, Valencia, Spain}
\author{F. Sanchez}
\affiliation{Institut de Fisica d'Altes Energies (IFAE), Bellaterra,
  Barcelona, Spain}
\author{M. J. Vicente Vacas}
\affiliation{Instituto de F\'isica Corpuscular (IFIC), Centro Mixto
  CSIC-Universidad de Valencia, Institutos de Investigaci\'on de
  Paterna, Apartado 22085, E-46071, Valencia, Spain}





\date{\today}

\begin{abstract}
We extend to 10 GeV results from a microscopic calculation 
of charged-current neutrino-nucleus reactions that do not produce a pion in the final
state.   
For the class of events coming from neutrino interactions with two
nucleons producing two holes (2p2h), limiting the calculation to
three-momentum transfers less than 1.2 GeV produces a two-dimensional distribution in
momentum and energy transfer that is roughly constant as a function of
energy.  The cross section for 2p2h interactions approximately scales with the
number of nucleons for isoscalar nuclei, similar to the quasi-elastic cross
section.  When limited to momentum transfers below 1.2 GeV, the cross
section 
is 26\% of the quasi-elastic cross section at 3 GeV, but
14\% if we neglect a $\Delta_{1232}$ resonance absorption component.
The same quantities are 33\% and 17\% for antineutrinos.
For the quasi-elastic interactions, the full nuclear model
with long range correlations produces an even larger, but
approximately constant distortion of the shape of
the four-momentum transfer at all energies above 2 GeV.
The 2p2h enhancement and long-range correlation distortions to the cross section for these
interactions are significant enough they should be observable in
precision experiments to measure neutrino oscillations and neutrino
interactions at these energies, but also balance out and produce less
total distortion than each effect does individually.

\end{abstract}

\maketitle


\section{\label{sec:introduction}Introduction}

Neutrino interactions in nuclei at energies up to 10 GeV are the core of current and
upcoming neutrino experiments to measure oscillation effects, 
neutrino interaction cross sections, and to search for new physics
beyond the standard model.  The precision of these experiments will be limited by
systematic uncertainties, likely including those from neutrino interaction modeling.
The on-axis NuMI flux has
modes that peak near 3 GeV or 6 GeV and serve MINOS and MINERvA with a
similar 3 GeV design proposed for LBNE.  The off-axis flux from the
same beam as used by NOvA peaks at 2 GeV and is tuned 
to include the energy of the expected maximum oscillation probability.
Even low energy experiments like T2K have a high energy tail, and
MicroBooNE will have a secondary peak at 2 GeV neutrino energy from
off-axis kaon decay neutrinos from the nearby NuMI beamline.

Measuring and modeling
neutrino-nucleus interactions has recently improved because of the
effort surrounding especially the
MiniBooNE double-differential quasi-elastic (QE) data \cite{AguilarArevalo:2010zc}. 
Those data peak near 600 MeV neutrino energy.
Several groups are investigating better models of the nuclear
environment described in the review article \cite{Morfin:2012kn}, especially the use of the random phase
approximation (RPA) to compute the effects of long-range nucleon-nucleon
correlations affecting the QE and $\Delta_{1232}$ interactions.
Computing the RPA series requires a model for the effective
(N and $\Delta_{1232}$) baryon-baryon interaction in the nuclear medium, including short-range
correlations (SRC).  
Also, a new class of interactions is now being computed where the reaction
involves two or three nucleons and producing two or three holes in the nucleus (2p2h and 3p3h).  
These components are required to describe
existing electron scattering data, and we find that they are also significant
for neutrino interactions \cite{Martini:2009uj, Martini:2010ex,
  Martini:2011wp, Nieves:2011pp, Nieves:2011yp, Nieves:2013fr,
  Amaro:2010sd, Amaro:2011aa}.

In this paper we present results from our microscopic model for charged-current
interactions that do not produce a pion in the final state,
now limited at 10 GeV and three-momentum transfer of 1.2 GeV,
the first detailed calculation of this type at these
energies.  
Previously computed results
\cite{Nieves:2011pp,Nieves:2011yp,Nieves:2013fr} 
for energies below 1.5 GeV
compare well with MiniBooNE data
and independent calculations by another group \cite{Martini:2009uj,Martini:2010ex}.
In this paper, these new calculations are also compared to other
predictions obtained empirically from electron scattering data and a
brief interpretation of existing data is included as well.

\section{\label{sec:structure}Structure of the calculation}

The high energy results presented here are the extension of a 
long-standing program to build up a complete microscopic calculation 
of the neutrino-nucleus cross
section \cite{Nieves:2004wx,Nieves:2005rq,Nieves:2011pp,Nieves:2012yz} 
which historically comes from work at neutrino energy around 150 MeV
\cite{valenciaSinghOset1998}.
\begin{figure*}
\includegraphics[width=5.8cm]{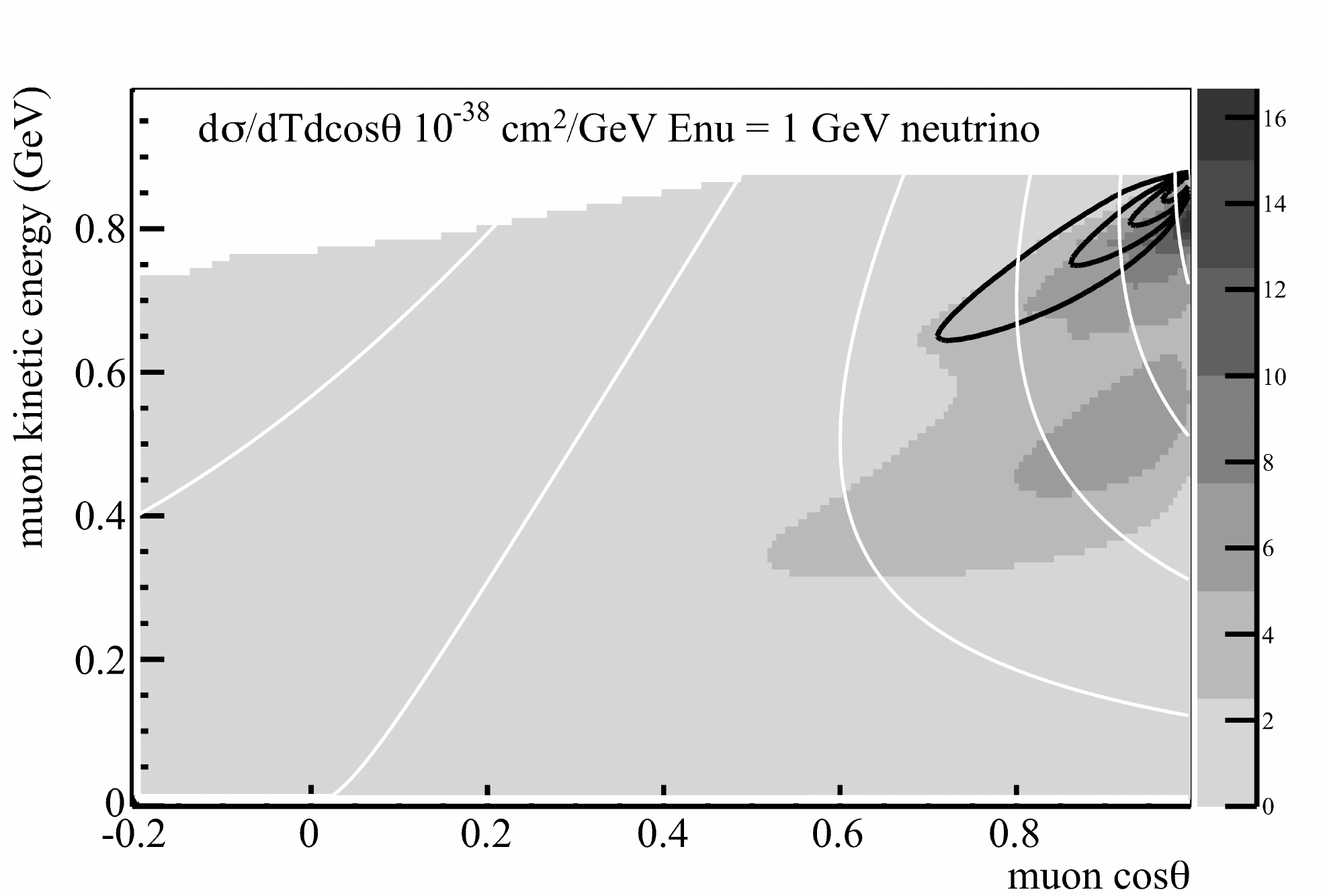}
\includegraphics[width=5.8cm]{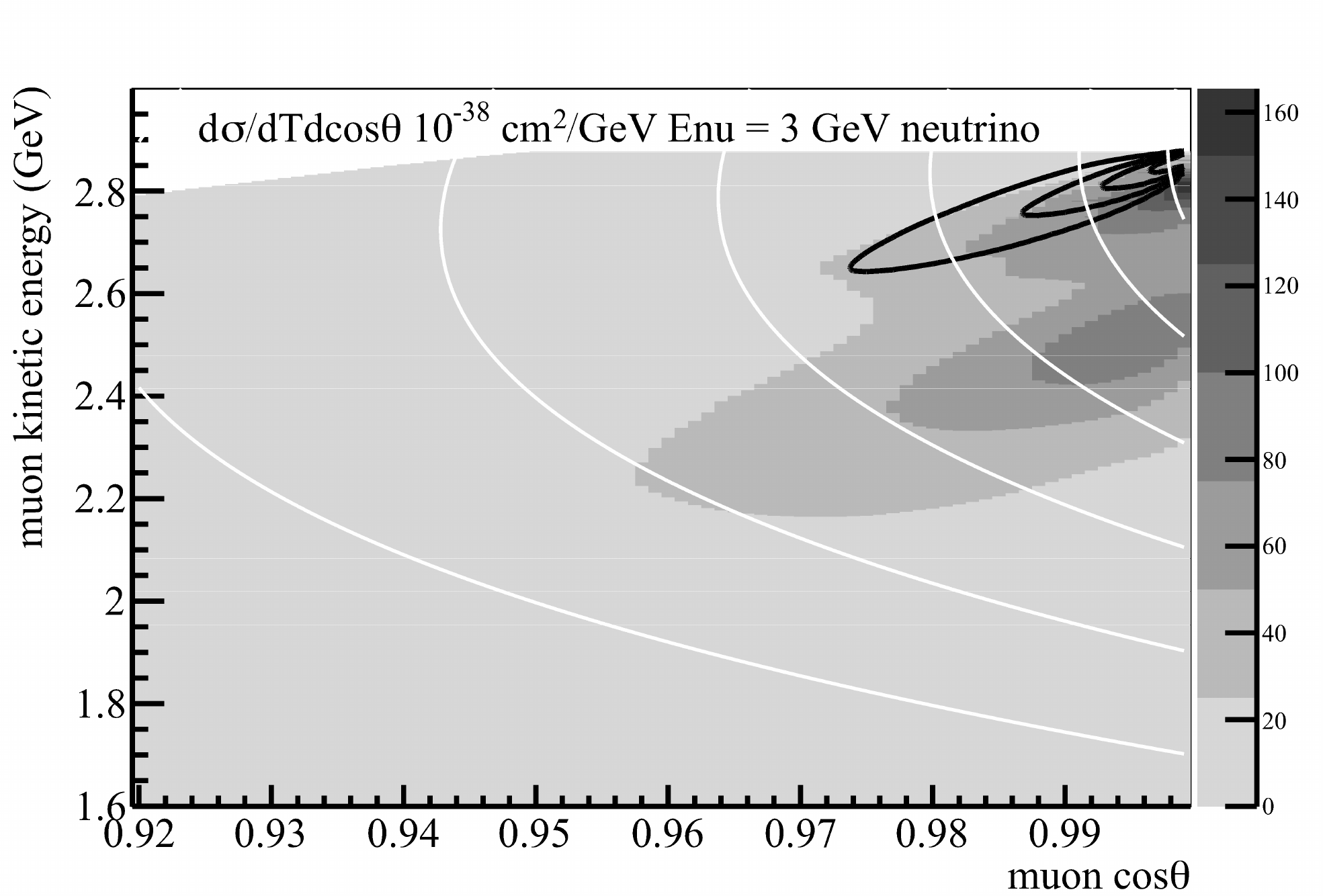}
\includegraphics[width=5.8cm]{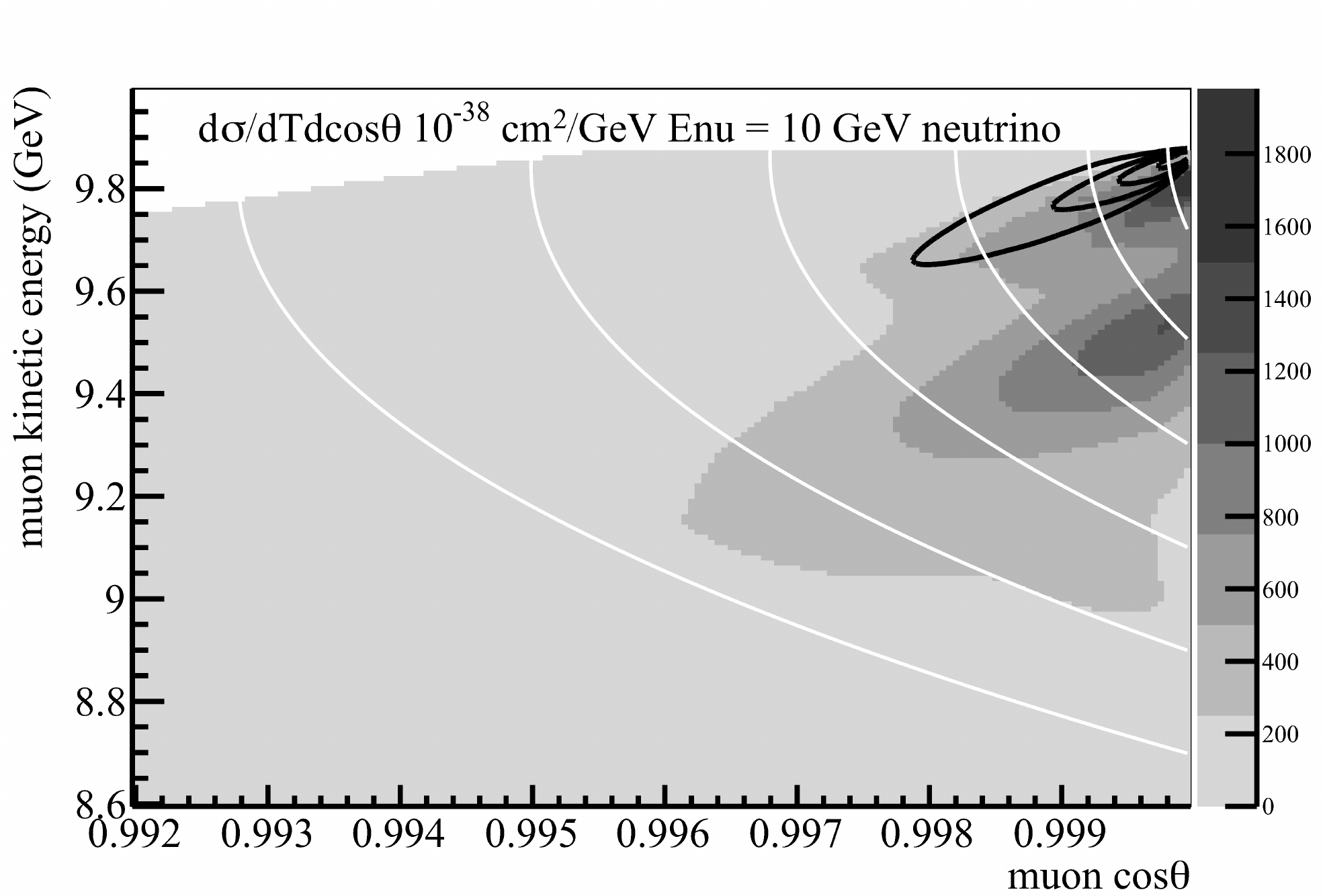}
\caption{\label{fig:neutrinoTC}  Double differential 2p2h cross section
  $d\sigma/dT_{\mu}dcos\theta_\mu$ ($10^{-38}$ cm$^2$/GeV) 
  for neutrino-carbon interactions at energies of 1.0, 3.0,
  and 10.0 GeV.  The black contours show the location of the QE events
  with four equally spaced contours from zero to maximum (which varies
  among the plots) and has a
  width due to the nuclear model.  The white shows lines of constant
  three-momentum transfer from 0.2 to 1.2 GeV.
}
\end{figure*}

The QE process uses a local Fermi gas (LFG) which includes Pauli blocking,
Fermi motion, removal energy, and Coulomb distortion.  Most significantly,
long and short range nucleon-nucleon correlations are included using the random phase
approximation (RPA) approach that accounts for both nucleon-hole and
$\Delta_{1232}$-hole components \cite{Nieves:2004wx}.  The free nucleon form
factors are the standard ones with an axial mass $M_A$ = 1.0 GeV
(different from \cite{Nieves:2011yp,valenciaSinghOset1998})
used in a dipole axial form factor and vector form factors from \cite{Galster:1971kv}.
In the results that follow we also show a comparison of these calculations to
the same free nucleon cross section applied within the local Fermi gas
nucleus but without RPA effects (noRPA)
which gives results within $\pm$ 5\% of the default QE cross section
of the {\small GENIE 2.8.0} neutrino event generator \cite{Andreopoulos:2009rq} used by
many experiments.  The differences can be attributed to the choice of
the axial mass parameter, vector form factors and the use of the
global Fermi gas in {\small GENIE} instead of a local Fermi gas in Ref.~\cite{Nieves:2004wx}.

The non-QE component presented here is constructed from a many body
expansion of modes where the exchanged W boson is absorbed by 
two or three nucleons. 
We refer to this set of
processes generically as two-particle two-hole channels (2p2h) which are
also called meson exchange currents.  The details are
described in \cite{Nieves:2011pp,Nieves:2011yp,Nieves:2013fr}
The equivalent
component for inclusive electron-nucleus scattering fills in the
so-called ``dip region'' between the QE and $\Delta$
peaks \cite{Gil:1997bm}, 
and plays the same role here in neutrino-nucleus scattering.  The calculation
includes processes that do not have a pion or an on-shell $\Delta$ in the diagram-level final
state, and so are ``QE-like'' by some experimental definitions.  

One class of 2p2h processes have $\Delta$ kinematics in which a
$\Delta$N$\rightarrow$NN absorption process occurs.  Some contemporary
event generators include pionless processes with $\Delta$ kinematics
using a $\Delta$ absorption process or a pion absorption with a 
final state interaction (FSI) cascade rescattering model, or both.
We can separate the pieces of the
calculation to not include the $\Delta$ absorption process, in
doing so approximately illustrate the size of the modification
relative to current neutrino interaction generator codes.  
This is done by subtracting the $\Delta$ absorption cross section, so
interference terms between the $\Delta$ resonance excitation mechanism
and non-$\Delta$ components are kept.

These calculations are made with no parameters tuned to neutrino-nucleus
data except for the choice of $M_A$ = 1.0 GeV for the axial form factor,
which is essentially tuned to deuterium bubble chamber data.

In previous work with neutrino energies below 1.5 GeV, the entire
kinematic space was well described by the model and its
calculations.  As the neutrino energy increases, it opens up a region
of high momentum and energy transfer in the kinematics.  The
model does not include 2p2h production via resonances beyond the
$\Delta$ or related interference terms.   Also, as the computation is
configured, the result is not adequately accurate for high
three-momentum transfers.  For both reasons, the 2p2h computation is
not suited to describe this high momentum-transfer region of kinematic space.

The low three-momentum transfer part of the calculation still remains as
accurate as it is at low neutrino energies, and includes most of the cross
section and the most interesting structure.
In addition, experimental analysis of charged current muon and
antimuon
samples with low hadron
multiplicity are often restricted to the forward direction due to
detector geometry design.  Higher momentum transfer events exit the
detectors out the side and are reconstructed with poor resolution or
cut completely.  The calculation we present here is well
suited to the most relevant energies and regions of kinematic
acceptance for current experiments.

\section{\label{sec:results}Results}

The three plots in Fig.~\ref{fig:neutrinoTC}
show the neutrino-carbon 2p2h cross section in the muon experimental observables
at energies of 1.0, 3.0, and 10.0 GeV. 
The bounds of each plot are constructed
so they contain events up to and a little beyond three-momentum transfer of 1.2 GeV.
The gray scale gives the 2p2h
double-differential cross section in units of 10$^{-38}$ cm$^2$/GeV.
Clearly evident is an upper
nondelta band and the lower $\Delta$
component.  Because we restrict the calculation to the nonresonant
and  $\Delta$ components, higher
resonance RN$\rightarrow$NN transitions (or their interference
effects)  do not appear below the delta band.
Two sets of contours are overlaid on the plot:  the black contours are
from the QE calculation; the white contours are lines of
constant three-momentum transfer up to 1.2 (in steps of 0.2) GeV.  
Lines of constant energy transfer can be inferred from the vertical axis.

As a function of energy, the structure and magnitude of the 2p2h cross
sections are quite stable.  In Table~\ref{tab:trend},%
\begin{table*}
\caption{\label{tab:trend}The 2p2h cross section in carbon vs. energy.
The contribution saturates as a function of three-momentum transfer
to a value that is 29\% of the QE cross section for neutrino, 32\% for
antineutrino, an estimate for the
nondelta component without the $\Delta$ absorption component
is 15\% and 17\% of the QE cross section for neutrino and
antineutrino. 
}
\begin{ruledtabular}
\begin{tabular}{l|cccccc}
  & \multicolumn{2}{c}{whole cross section (x 10$^{-38}$ cm$^2$)} & \multicolumn{4}{c}{ three-momentum transfer $<$ 1.2 GeV}\\
Energy &  QE  &  QE  & 2p2h & 2p2h & QE & QE \\
(GeV) & LFG+RPA & LFG noRPA & & no $\Delta$ & LFG+RPA & LFG noRPA\\
\hline 
1 $\nu_\mu$ & 5.61 & 5.66 & 1.27 & 0.563 & 5.20 & 5.36 \\ 
2 & 5.65 & 5.61 & 1.41 & 0.704 & 4.52 & 4.74\\
3 & 5.45 & 5.45 & 1.43 & 0.735 & 4.30 & 4.54\\
5 &  5.22 & 5.25 & 1.46 & 0.761 & 4.14 & 4.39\\
10& 5.04 & 5.10 & 1.47 & 0.781 & 4.01 & 4.27\\
\hline
1 $\overline{\nu_\mu}$  & 1.56 & 1.96 & 0.459 & 0.306   & 1.56   & 1.95  \\
2 & 2.68 & 3.03 & 0.887 & 0.520 & 2.52 & 2.89\\
3 & 3.26 & 3.55 & 1.07 & 0.609 & 2.93 & 3.27\\
5 & 3.83 & 4.05 & 1.24 & 0.686 & 3.29 & 3.61\\
10 & 4.31 & 4.47 & 1.38 & 0.749 & 3.58 & 3.88\\
\end{tabular}
\end{ruledtabular}
\end{table*}
we show the integral
of the QE and 2p2h cross sections within the three-momentum transfer
$q_3 < 1.2$ GeV contour.  
In addition, the total QE cross section is
also given for carbon.  
The computed QE cross section
decreases slowly with energy following the inherent dependence of the
free-nucleon cross section.  
The 2p2h estimates are slowly increasing primarily because the
calculation is adding more cross section below the $\Delta$ near the
right axis.  With the $\Delta$ component, the 2p2h cross section is
26\% of the total QE cross section, without the $\Delta$ absorption
component it is 14\% for 3.0 GeV neutrino interactions, and rises to
32\% and 17\% at 10 GeV respectively.

\subsection{Momentum and energy transfer 2D plane}

\begin{figure}[hbt!]
\includegraphics[width=8.8cm]{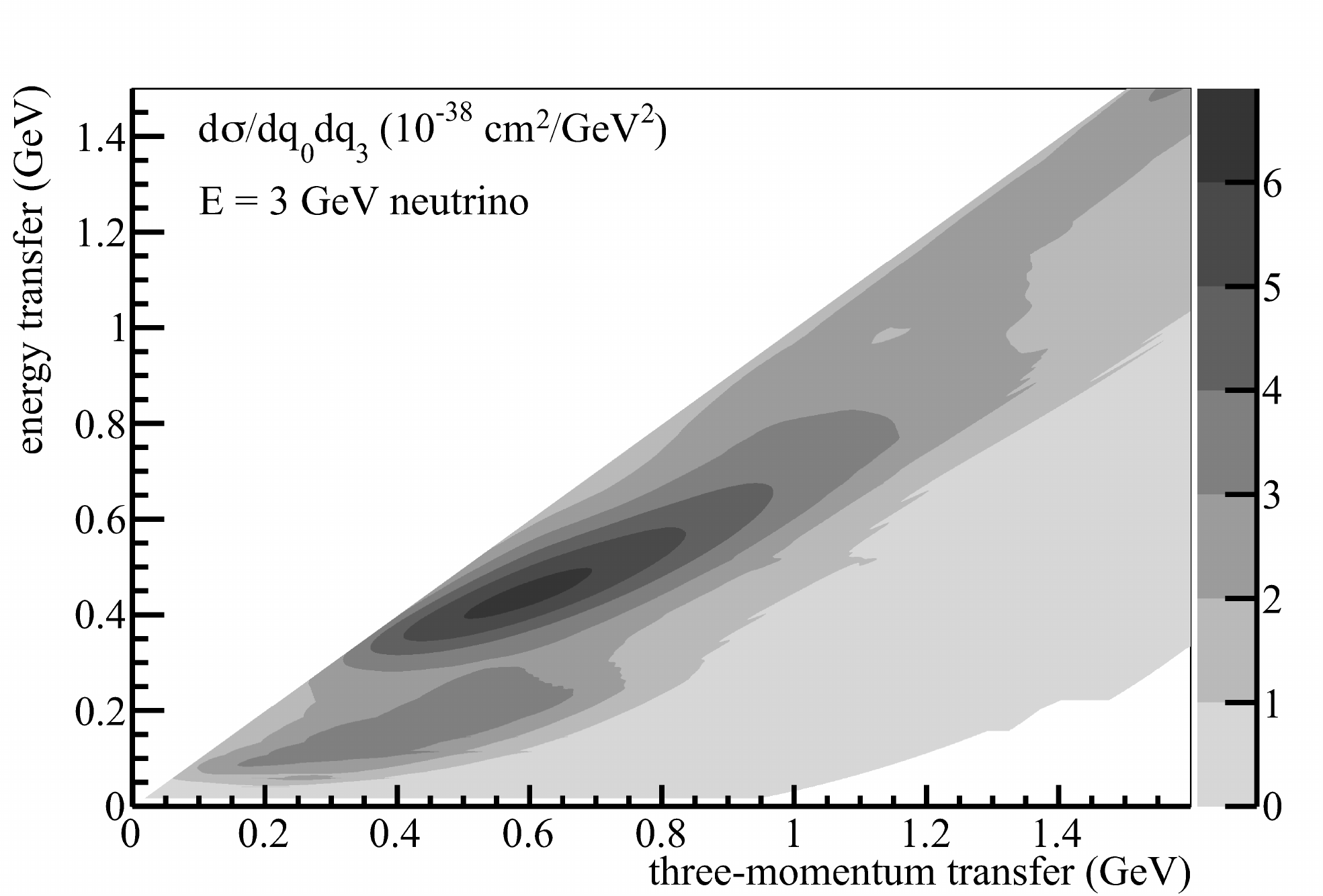}
\includegraphics[width=8.8cm]{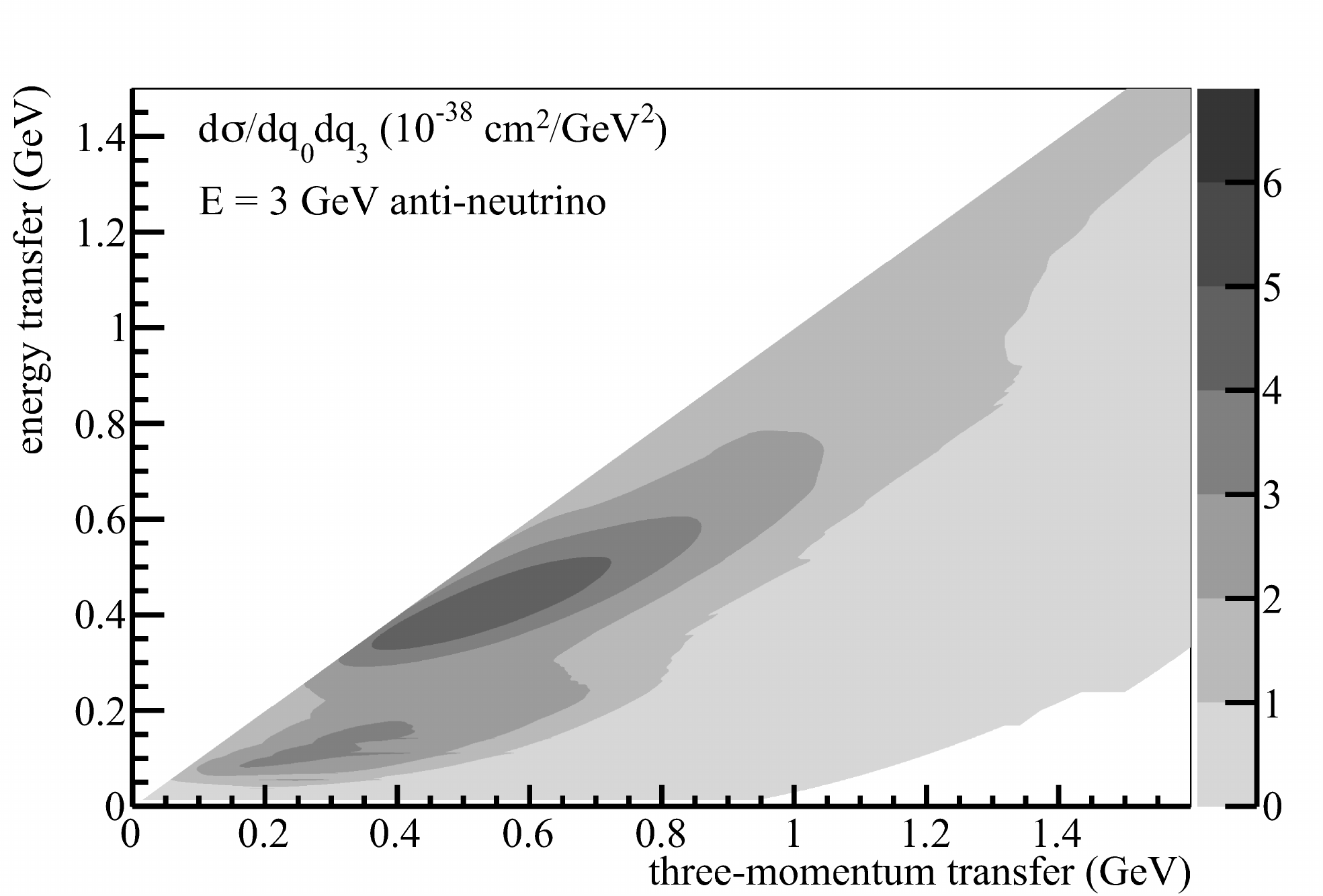}
\caption{\label{fig:q0q3} The 2p2h cross section
  $d\sigma/dq_0dq_3$  
  vs. energy transfer and
  three-momentum transfer for 3.0 GeV neutrinos (top) and
  antineutrinos (bottom).
}
\end{figure}

These cross sections are naturally better expressed in terms of 
momentum and energy transfer,
so the three plots shown above
can be summarized as in Fig.~\ref{fig:q0q3}.  In these
kinematics, the $\Delta$ component is the top peak, and the
non-$\Delta$ part peaks lower, just above the QE kinematics
and  fills in the
dip region. As mentioned above, the $\Delta$ component peak could be
mostly assimilated to the $\Delta$N$\rightarrow$NN absorption
process; it is the production of an on-shell $\Delta$, subsequently reabsorbed by 
another nucleon. Such contribution is often not  included in what is 
commonly called a meson-exchange current.
In the energy range from 2.0 to 10.0 GeV, the cross section in and
surrounding the two peaks stays within 10\% of the value shown here, 
because the hadronic tensor does not change and the leptonic part of
the calculation changes slowly.
In the tails of the distribution, especially along diagonal 
$q_0 = q_3$  
edge, the differences grow to 30\% with higher energy.

Both parts of the 2p2h cross section appear in a different location
than the QE part of the cross section.  Rather than being near 
$W_{\Delta} = 1.232$ GeV or nucleon $W \approx 0.938$ MeV, the non-$\Delta$ component
peaks near a line of $W \approx 1.00$ GeV ($W^2 = M_N^2 + 2 M_N q_0 +
q^2$) at very-low $-q^2 = Q^2 < 0.5$ GeV$^2$ 
with a substantial asymmetric tail toward the $\Delta$ and higher energy
transfer.  At higher $Q^2$ the 2p2h peak crosses under the QE line but
retains the asymmetric tail.  In all cases, the 2p2h contribution is
wider than the QE and effectively fills in the QE and $\Delta$ and the
dip region between them.  
The $\Delta$ absorption component peaks at $W = 1.232$ GeV as expected.

An experiment
that classifies these as QE-like, because no pion was observed and/or
because the proton was below the reconstruction threshold, might
choose to use the QE lepton kinematics to reconstruct the neutrino
energy.  See, for example, the discussion in \cite{Nieves:2012yz}.
Though the QE events will be unbiased up to the average removal
energy estimate, most of the 2p2h nondelta component will pick up a bias
which is typically 100 MeV below the true
neutrino energy while the $\Delta$ component will be centered 350 MeV
low.  These estimates are constant with neutrino energy, so they
become a smaller fractional bias as neutrino energy increases.
Likewise, if the biased energy estimate is then used to make an
estimate of the reconstructed $Q^2$, that too will be biased low.

{\bf Antineutrino case}  All the trends for the antineutrino case are
similar to the neutrino case, and are included in Fig.~\ref{fig:q0q3} and
Table~\ref{tab:trend}.  The 2p2h components of the antineutrino case
rise similar to the underlying QE antineutrino cross section and are 33\% and 19\% with and
without the $\Delta$ absorption component at 3 GeV, relative to the
QE+RPA cross section.  This is a somewhat higher fraction relative to
the QE rate than the neutrino version, and also the QE with and
without RPA are themselves 9\% different at 3 GeV, converging as
energy rises.

{\bf Application to event generators}  The distinction between the 2p2h
cross sections with and without the $\Delta$ component is important.
A portion of the cross section involving a $\Delta$, corresponding
specifically to $\Delta$ absorption, can be incorporated into a modern event generator via its
treatment of $\Delta$ and/or pion final state reinteractions in the nucleus.  Simply adding
the full cross section presented here could double-count some of these
events, so Table~\ref{tab:trend} also gives the cross section without $\Delta$ absorption.
The alternative approximation is to discard events from an event generator where a
$\Delta$ was absorbed and keep the whole pionless cross section estimate
described here. 

{\bf Uncertainty on the calculations}  Though the prediction for the cross section within a choice of
three-momentum cutoff is stable, and the differential cross section
itself is small, a substantial amount of 
cross section is not included in the integration because of the large kinematic space.
Moving the cutoff value back to 1.1 GeV or forward to 1.3 GeV reduces
or increases the integrated cross section by about 10\%, or about 8\%
for the component without the $\Delta$.

The above variations occur without including higher
resonances. As the neutrino energy increases, the second resonance
region of nucleon excitations ($P_{11}(1440)$, $D_{13}(1520)$ and
$S_{11}(1535)$), followed by absorption on another nucleon, 
might play some role at high energy transfers, above
$q_0 >0.9-1$ GeV (note that $q^2< 0$ and hence $q_3$ is always
bigger than $q_0$).  According to Refs.~\cite{Leitner:2008ue,
Leitner:2009zz}, the $N(1520)$ resonance would be the only one whose
effects might not be totally negligible, at least for neutrino energies
below 2\,GeV. But even this latter $N^*$ contribution represents
only a quite small fraction of that of the $\Delta$. The results
shown in Fig. 5.11 of \cite{Leitner:2009zz} illustrate this especially
well.  There, the integrated cross section for CC induced
resonance production on the proton and on the neutron are
displayed. Nevertheless, this source of uncertainty might contribute
an additional 10\% to the high momentum transfer part of the
calculation.   However,
the cross section involving higher resonances (neglecting interference contributions)
is already
incorporated into most of the modern event generators, as
mentioned in the paragraph above for the case of the $\Delta$, the
issue is how much migrates to a QE-like final state because of
in-medium absorption of the resonance state.  Because
of the dominance of the $\Delta$, and the important modifications of
its properties inside of a nuclear medium, we have prioritized
its contribution,  together with that of the non-resonant
background terms {\em and} the quantum mechanical interferences among all of
possible mechanisms, a combination that is usually not considered in the 
neutrino interaction event generators. 

\subsection{\label{Q2}Four momentum transfer distributions}

\begin{figure}
\includegraphics[width=4.3cm]{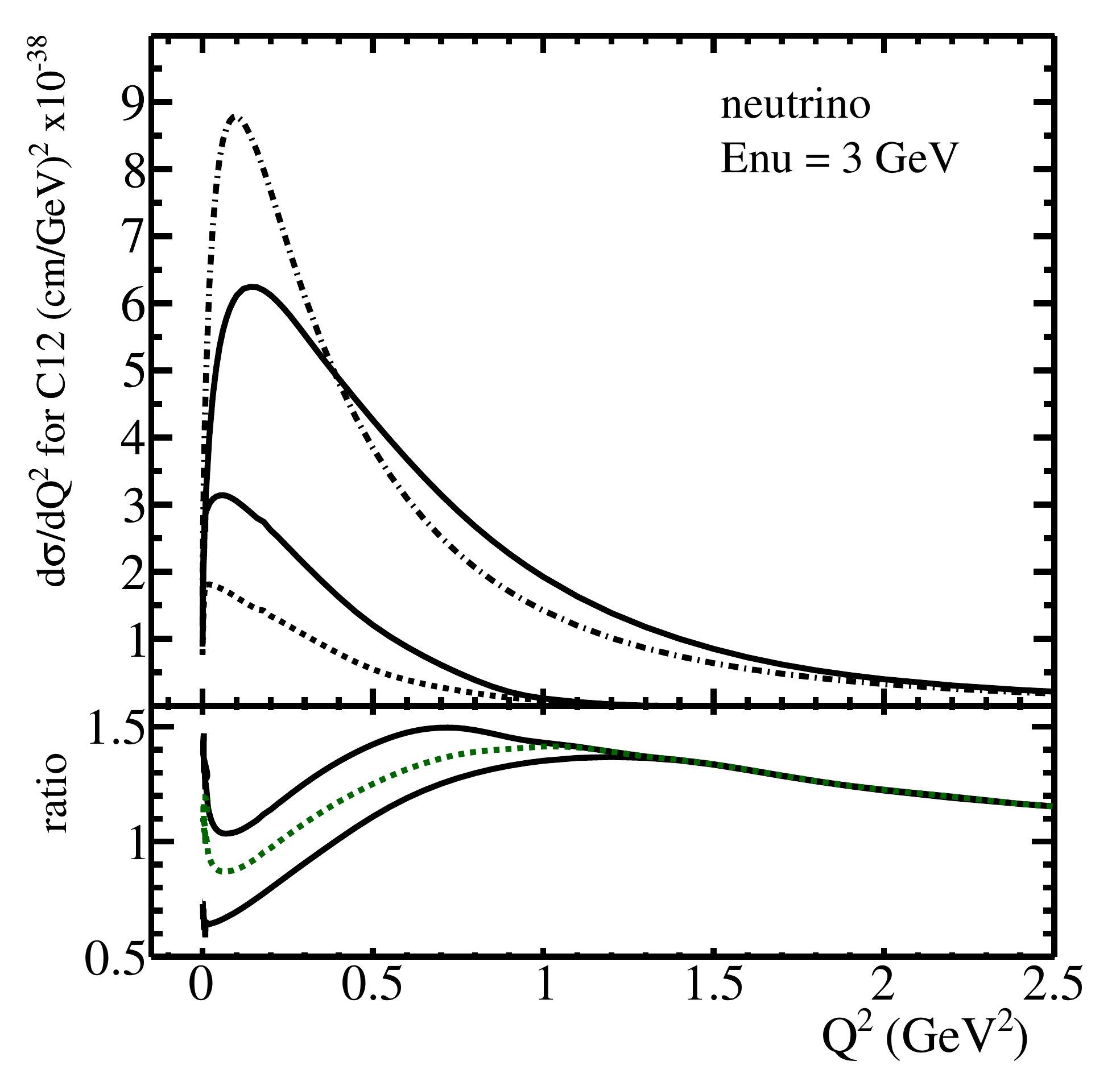}
\includegraphics[width=4.3cm]{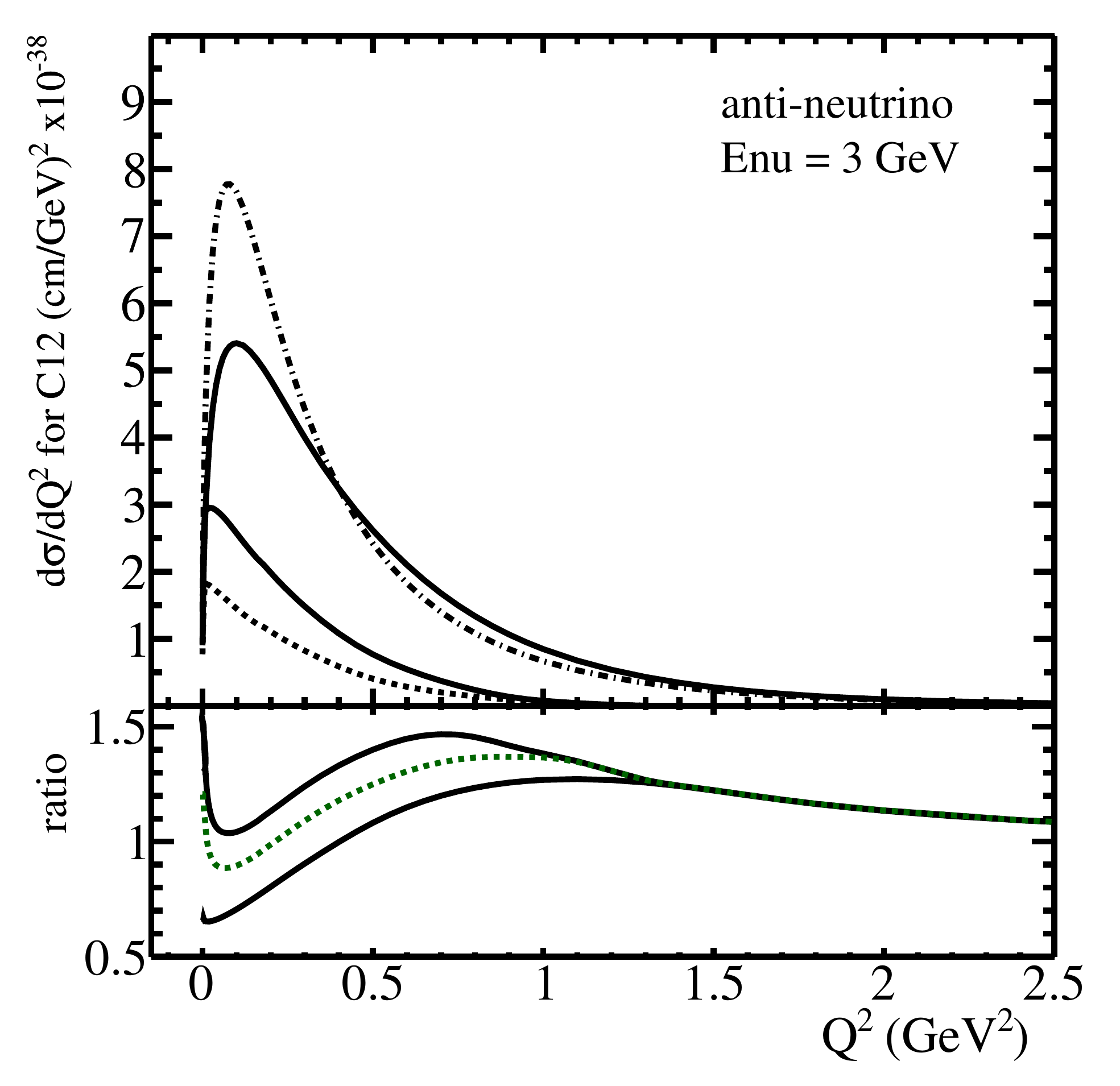}
\caption{\label{fig:qq} The $Q^2$ distributions for QE and 2p2h
  contributions, neutrino case (left) and antineutrino (right) for
  incoming energy of 3 GeV.  The QE and 2p2h cross sections are the solid upper and solid lower
  lines respectively.  Upper dashed line is the QE without RPA; lower dashed line is the 2p2h cross section
  without the $\Delta$ absorption component.  
  The lower solid ratio line is QE$_{\mathrm{full}}$/QE$_{\mathrm{noRPA}}$,
  the dashed ratio line is
  (QE$_{\mathrm{full}}$~+~2p2h$_{\mathrm{no\Delta}}$)/QE$_{\mathrm{noRPA}}$,
  and the upper ratio line is (QE$_{\mathrm{full}}$~+~2p2h$_{\mathrm{with\Delta}}$)/QE$_{\mathrm{noRPA}}$.  For all
  variations, the QE lines are the complete cross section but the 2p2h
  lines truncate the integration at 
  $q_3 = 1.2$ 
  GeV, causing it to
  also not contribute to the ratio above $Q^2 = 1.2$.
}
\end{figure}

Integrating the cross section in the previous figures along
lines of $Q^2$ gives Fig.~\ref{fig:qq}. %
The solid lines are from the calculation described above; the top solid one 
is the full QE model with local Fermi gas and nucleon correlation effects,
the bottom one is the full 2p2h contribution.  
The dashed lines are special versions for comparison.  
The top is QE without RPA, similar to
the standard treatment for neutrino experiments.
The bottom dashed line is the 2p2h contribution without
the delta absorption component.
The QE components show the full cross section, integrated all the way
to the end of the appropriate $Q^2$ contour.  The integration for the 2p2h contribution is stopped at the
three-momentum 
$q_3 = 1.2$ 
GeV boundary, like the values in
Table~\ref{tab:trend}.  

The bottom portion of each figure shows two
ratios.  The lower solid curve is the ratio of the full QE model to QE
without RPA.  The dashed curve is the ratio of
( QE$_{\mathrm{full}}$+ 2p2h$_{\mathrm{no\Delta}}$) /
QE$_{\mathrm{noRPA}}$, and the top solid curve is like the dashed
curve but with the 2p2h $\Delta$ absorption component.

The 2p2h model contributes events at lower $Q^2$, and
especially modifies the total for $Q^2 < 0.2$ GeV$^2$ where the QE rate is
reduced due to Pauli blocking and RPA effects.  
Through the middle of the $Q^2$ region,
it causes a mild shape distortion.  
For comparison to experimental results, the reconstructed $Q^2$ distribution will be
further distorted, a little bit if calorimetry is available to
estimate the neutrino energy, and a larger amount if the QE assumption and
lepton kinematics are used.  For the neutrino case, 2p2h interactions will be
reconstructed low due a biased low E$_\nu$ estimate.
At 3 GeV, this migration causes the reco version of the 2p2h cross
section in Fig.~\ref{fig:qq} would be 20\% higher near $Q^2 = 0$ and
20\% lower near $Q^2 = 1$ GeV$^2$; at 10 GeV the effect is one-quarter
this size.
The bias due to calorimetry or two-particle kinematic reconstruction
is more difficult to assess without
explicit final state nucleons and final state reinteractions, which
are beyond the scope of this paper. If they are put in a sample and
reconstructed as if they are QE, the expectation is neutrino QE has
energy deposits from a proton in the final state while 2p2h have a mix
of $pn$ and $pp$, so the 2p2h component will have more missing energy.
For antineutrino events, the 2p2h component mix of
$pn$ and $nn$ will more often appear to have protonlike energy
than expected for a pure QE sample with its neutron final state.

For experiments that are sensitive to the shape of the $Q^2$ distribution
of a QE-like signal, the inclusion of nucleon-nucleon
correlation effects in the RPA series yields a much larger shape distortion toward relatively more high-$Q^2$
interactions, with the 2p2h component filling in the suppression at
very low $Q^2$.   
Correlation effects in this model are tuned to low energy nuclear
phenomena, such as electron scattering and muon capture on nuclei, where they are essential
for a good description of data \cite{Nieves:2004wx}.  The suppression to a factor of
0.6 at $Q^2 = 0$ is the same kinematics, and is the most robust part
of this calculation.  The point near $Q^2 = 0.4$ GeV$^2$ where the effect
changes from suppression to enhancement is also where
the tuning of correlation effects is well constrained \cite{Gil:1997bm}. 
In the calculation, the RPA effects go to 1.0
at very large $Q^2$ because sizes larger than one nucleon are no
longer being probed.  Technically, there
should be a transition to probing neutrino-quark scattering which is
not part of this calculation.

The low $Q^2$ suppression is a combination of both short and long
range correlation effects.  The trend moving toward $Q^2 = 1.1$
GeV$^2$ is an
enhancement of the cross section but leaves the region where the
model was tuned to other nuclear effect data, and this specific part
is not relativistic, hence the model suffers from larger uncertainties.
The maximum enhancement is 35\% for the
neutrino case and 25\% for antineutrinos within our approach.
These numbers should be taken with caution, since the model has been
extracted beyond its reliable range.  An alternate version of
the calculation which has a covariant form and a 20\% lower cross section
at 1.1 GeV$^2$ roughly indicates the size of the uncertainty, though
should not be considered a one-sigma statement.  It is reasonable to
expect some enhancement due to RPA above $Q^2 = 0.5$ GeV$^2$, before RPA
effects become negligible.
This feature is driven
toward enhancement by the transverse part of the ph-ph interaction
(second term in Eq. (36) of \cite{Nieves:2004wx})
when it and also the longitudinal term both change sign.  
As the momentum transfer increases, the non-relativistic form of the
first term in brackets and the
simplification of g'(q) = g' = 0.63, which neglects the mild q
dependence, contribute to uncertainty in the
maximum size of the enhancement and how fast it drops to zero.

This higher $Q^2$ region is where the short range correlation (SRC) effects
are most important.  The model parameters
are not specifically tuned to the equivalent electron scattering data,
which also show an excess of cross section strength when SRC are not
accounted for.  A substantial SRC component in electron
scattering is needed to reproduce for the tail of the measured nucleon
momentum distribution and an enhancement of the cross section in
kinematic regions away from where 2p2h and FSI contributions play a
significant role.
A recent and
comprehensive review in electron scattering is provided by \cite{Arrington:2011xs}.
Another review \cite{Hen:2013oha} covers the SRC portion but also
emphasizes a phenomenological connection with
the EMC effect in deep inelastic scattering.
Because we expect the 2p2h contribution to be small at these values of
$Q^2$, neutrino scattering data with excellent coverage of
the $Q^2$ = 1 GeV$^2$ region may also be an interesting new window to
understanding this feature of the nuclear environment, 
and we include some discussion in a later section.

Not shown in these plots, the distortion as a
function of $Q^2$ for all energies above 2 GeV is essentially constant.
Compared to the 3 GeV calculation shown, it remains within 5\% at all
$Q^2$ away from the backscatter limit.
The antineutrino case is
similar, though the enhancement at high $Q^2$ is slightly less
pronounced and as shown in Table~\ref{tab:trend} the resulting genuine
cross section remains around 4\% lower than the model without RPA, even at 10 GeV.

\subsection{Isospin content of the initial state} 

\begin{figure}[hbt!]
\includegraphics[width=8.8cm]{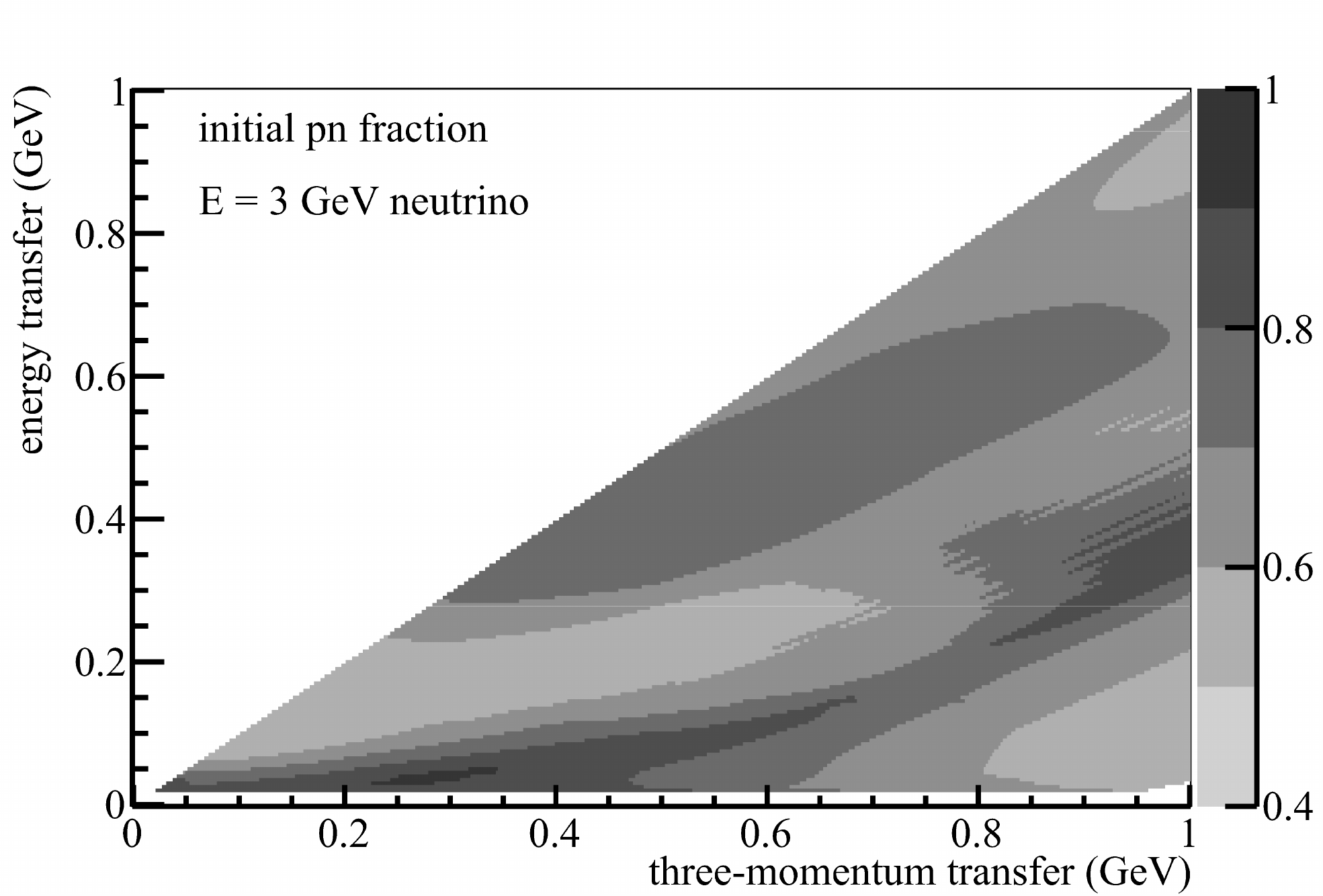}
\caption{\label{fig:isospin} The fraction of the 2p2h cross section
 coming from an initial $pn$ pair for 3 GeV neutrinos.
The antineutrino trends are very similar.  
The momentum transfer axes only go to 1.0 GeV in this plot. 
}
\end{figure}

The 2p2h calculation yields 67\% of the cross section coming from $pn$ pairs
in the nucleus, for neutrino energy of 3 GeV, when the cross
section is integrated to 
$q_3 = 1.2$  
GeV.
Part of this is from the $\Delta$ absorption component which
is explicitly given an initial state $pn$ fraction of 5/6.
The portion of the 2p2h cross section not from $\Delta$ absorption
(including the interference term) has only a 50\% fraction coming from an initial $pn$ state.
These results hold for the charged-current neutrino case
$W^{+} + np \rightarrow pp$ shown in Fig.~\ref{fig:isospin} and also for the  antineutrino $W^{-} + np \rightarrow
nn$.
The isospin content is not uniform in the kinematic plane.  In
addition to the 5/6 $pn$ initial state at $\Delta$ kinematics and
50\% $pn$ initial state at the non-$\Delta$ peak.  There is a ridge of high
$pn$ initial state just below this from the interference term and
extending underneath the location of the QE peak.

Electron scattering measurements of SRC effects summarized in \cite{Arrington:2011xs}
show the initial state for the SRC process is $>90$\% $pn$ pairs,
deduced from different types of measurements.
Our model for QE does not provide a
prediction for this aspect of the process.  
Given the charged-current nature of the interaction, a reasonable
guess is the neutrino case (before
hadron rescattering) would have an excess of outgoing $pp$ relative to
$pn$ in which 
the supposed spectator
nucleon shows a large momentum opposite to the initial state of its
struck partner, with the antineutrino providing the same for nn
pairs.  This would be a different character than the low $Q^2$ 2p2h
estimate presented here, though it is similar to the portion of of the
2p2h cross section with QE kinematics affected by the interference
with the $\Delta$ absorption component.
Overall, this model predicts a complicated isospin
dependence that would vary substantially with different lepton kinematics.

\subsection{Variation with the size of the nucleus}

The 2p2h cross section without the $\Delta$ absorption component, integrated to
$q_3 = 1.2$ 
GeV,
varies linearly with the size of an isoscalar nucleus.
The 2p2h cross section including $\Delta$ absorption shows
deviations from linear dependence; $^{16}O/^{12}C$ = 1.5 and $^{40}Ca/^{12}C$=4.0,
though the $\Delta$ component is expected to be nonlinear in this
way.
Thus, the full neutrino 2p2h cross section grows a bit
faster than the number of nucleons, behavior that looks compatible with the
results observed for electron scattering \cite{VanOrden:1980tg,DePace:2004cr}.

Typical also of QE calculations with realistic nuclear models, as the
nucleus size increases, the
cross sections are lower at forward angles and very-low $Q^2$ and
enhanced at very low energy transfer.

\section{\label{discussion}Discussion}

There are experimental data and two other models in this energy range
available for initial comparisons.

\subsection{\label{susa}Super scaling approximation (SuSA)}

Super scaling approximation model is a fully relativistic scheme \cite{Amaro:2004bs}
that provides a good representation of all existing QE electron
scattering data for high enough momentum and energy transfers, to
the extent that quasi-elastic scattering can be isolated. Thus,
it incorporates the correct $q_3$ and $q_0$ dependence of the
$(e,e')$ spectrum.  The SuSA model is expected to provide a good 
description of $\nu$ and $\bar \nu$ CCQE data lacking only the
two-body 2p2h contributions which will increase the QE-like cross
sections. It has been recently extended up to very high neutrino
energies \cite{Amaro:2013yna} and compared to MiniBooNE and NOMAD CCQE
cross sections.  The SuSA model predicts cross sections
that saturate with neutrino energy, like the underlying QE process, and
are always lower than a Fermi gas model, the same behavior as the RPA
model presented here.  On the other
hand, SuSA underestimates the MiniBooNE data, leaving enough room 
at these energies for the 2p2h contributions studied in this work.

Our RPA model was tuned to low momentum transfer data, including muon
capture, and the SuSA expression of moderate momentum transfer
electron scattering data is complementary.  Though comparisons
to just the $\sigma(E)$ cross sections are indirect, the results
in Figs. 1 and 2 of \cite{Amaro:2013yna}, both absolute and relative
to the Fermi gas model they present, compare well with our RPA model.
The SuSA predictions fall between our default 
model and an artificial estimate with zero enhancement at moderate and
high $q^2$.  This observation 
suggests that the intermediate, explicitly relativistic
alternate version in our paper is close to what an extrapolation of
moderate $q^2$ electron scattering data would prescribe.

\subsection{\label{tem}Transverse enhancement model (TEM)}
An empirical extraction of missing components of the cross section
has been obtained from electron scattering data, along with a suggestion for how to
approximate it in the neutrino case \cite{Bodek:2011ps}, which they refer
to an enhancement of the transverse component of the cross section.
This extraction of the cross section was done with inclusive electron
scattering and a model that included $\Delta$ production.  Under the
assumption that the enhancement is coming from the 2p2h component
and/or long and short-range correlations, the appropriate comparison
is the version of our model which
does not include the $\Delta$ absorption component.

In \cite{Bodek:2011ps},
application of the transverse enhancement to the neutrino case is to
modify the $G_{Mp}$ and $G_{Mn}$ form factors as the same function of $Q^2$
that described the electron scattering data, and not change the
longitudinal or axial form factors.
As implemented, this insight is a function of $Q^2$ only, and does not
preserve the kinematic features that fill in the dip of the electron
scattering data from which it was obtained.  Despite this, a
prediction for the distortion of the $Q^2$ distribution is presented
for a neutrino energy of 3 GeV.

At $Q^2 = 0.2, 0.5, 1.0, 2.0$  GeV$^2$ it enhances the cross section by
20\%, 30\%, 18\%, and 5\% , but the enhancement approaches zero as
$Q^2$ approaches zero.  Their enhancement also is constant with energy for the QE
process above 2.0 GeV neutrino energy.
Our 2p2h without $\Delta$ absorption model yields an enhancement of 18\% and 15\% for the first
two data points, and in general the contribution to the
cross section continues to rise from $Q^2 = 0.5$ GeV$^2$ down to
zero.   Because we have truncated the 2p2h model, 
the values and trend in the higher $Q^2$ regions cannot be compared.
However, the RPA enhancement is also driven by the transverse
component, and in combination with the 2p2h model might be describing
the same underlying physics.  In this case the magnitude and direction
of the enhancement is similar, with the TEM reaching its maximum
enhancement earlier in $Q^2$ than the model presented in this paper. 

\subsection{\label{nomad}NOMAD}

The NOMAD experiment analyzed a set of QE-like interactions on
carbon \cite{Lyubushkin:2008pe} whose flux has an average energy of
25.9 GeV for neutrino and 17.6 GeV for antineutrino.  The average
energy is high due to a long high energy tail.  The total neutrino
event rate peaks near 5 GeV, so there is a substantial portion of
their event rate between 3 and 10 GeV.

They divide their data into a two-track sample which is primarily $Q^2$
above 0.3 GeV$^2$ (Fig. 14 in their paper) and also a one track sample
which is primarily low $Q^2$ but includes higher $Q^2$ interactions
where the proton was not reconstructed.  From this they estimate how
much they should enhance the QE cross section so that their simulation
describes the data, and also how to modify the axial mass parameter so
the simulation describes the $Q^2$ shape of the data.

Our model produces a low $Q^2$ sample that is suppressed by the RPA effects,
but some of the cross section returns with the addition of the 2p2h
contribution.  Quantitatively how strong this is for a NOMAD-like one-track
sample depends on what fraction of the sample comes from higher $Q^2$
interactions, which is not provided.
The high $Q^2$ sample is made with a selection that requires the
kinematics to agree with the QE prediction with little missing momentum.
This should systematically reject a large fraction of the 2p2h
component, as well as QE and pion production where the hadrons
rescattered as they exited the nucleus.  A sample like this could be
a very good place to test the effects of SRC alone.  Our model
predicts an overall enhancement of the cross section of around 15\%
and that RPA effects would give a relative deficit at 
$Q^2$ = 0.3 and excess at 1.5 GeV$^2$ compared to QE without RPA.

In the NOMAD analysis, a large source of
uncertainty comes from their final state interaction model, which is
implemented within the package DPMJET \cite{Ranft:1988kc}, 
a calculation 
developed for TeV accelerator and cosmic ray modeling of hadronic shower development.
The NOMAD authors tune a "formation time" parameter $\tau_0$ to
the data without considering RPA or 2p2h effects, common procedure in
that era.  They present results repeating their analysis with
three different amounts of final state interactions, to illustrate the
model agreement to the data regardless of the tuning.  With more
final state interactions (by decreasing $\tau_0$)
the trend is to need fewer events in the one-track sample and more
events in the two track sample.  The lowest parameter value they
tested, $\tau_0 = 0.6$ (more FSI than their favorite tune) is close to
the 0.5 value they determined from their tuning procedure to be an
appropriate one-sigma extreme.  For this choice of parameter, their simulation
underpredicts the two-track event rate by 8\% but has the low-$Q^2$ one-track
event rate about right.
The shape fit returned a poor chis-quare and they do not show the distributions, 
but the other two fits were trending toward
the shape distortion we describe, and their Fig. 14 , with its $\tau_0$
parameter at 1.0,  already shows a mild distortion in the $Q^2$ shape 
which is just under half of what our model suggests. 
A quantitative analysis cannot be done
without more information about the acceptance and FSI model, and our
SRC part of the model has a significant uncertainty in the $Q^2$ region of their
two-track sample. However, the range of results within the context of
their analysis certainly allows for the presence of substantial RPA
and 2p2h effects in the data.

\subsection{\label{sec:otherexperiments}Other data from $Q^2$ shape
fits}

Other high statistics experiments with substantial event rate above 1
GeV include the published result from the K2K SciFi
detector \cite{Gran:2006jn}
and results from
K2K SciBar \cite{Espinal:2007} and MINOS \cite{Dorman:2009} in
conference proceedings.
They report high fit values when extracting an
axial mass $M_A$ parameter using a shape-only fit to the $Q^2$
distribution.
The fundamental observation is the simulation overpredicts the relative event rate at
very low $Q^2$ and underpredicts the rate at high $Q^2$.  Again, effects due
to RPA and 2p2h were not routinely considered at the time of those
analyses, 
so the suggestion is a combination of high $M_A$ and additional
suppression at very-low $Q^2$ could describe the data.
The model presented here, with RPA providing a large $Q^2$ shape
distortion and modest 2p2h contribution returning some but not all
event rate at very-low $Q^2$ has the same features.

\subsection{\label{sec:miniboone}MiniBooNE and SRC}

Most of the MiniBooNE data is at lower energy than we consider
in this paper and comparisons with this model have already been made
\cite{Nieves:2011yp}, but the effects on the $Q^2$ distribution due to RPA 
are the same
until close to the backscattering kinematic cutoff. 
Data from the higher energy portion of the MiniBooNE flux
produces significant event rate with $0.5 < Q^2 < 1.0$ GeV$^2$, which is
where our model predicts a relatively small enhancement of the cross section due
to 2p2h events but a significant enhancement from the transverse part
which strongly depends on the SRC part of the RPA model.
Though the SRC model is not tuned to neutrino data,
the enhancement does contribute to the agreement at high
$Q^2$ (and high $E_\nu$) portion of the data.  

The
calculation for the MiniBooNE flux is shown in Fig. 3 of
Ref. \cite{Nieves:2011yp} with and without RPA
effects for one angle bin.   Most of the discussion in
\cite{Nieves:2011yp} 
focuses on how well the combined low $Q^2$ RPA and 2p2h contributions describe the
MiniBooNE double differential cross sections.  In the context of the
results presented here, we call attention to regions of the cross section where
the SRC effects are particularly significant.
The $T_\mu$ range from 1 GeV to 1.5 GeV in that figure
corresponds to this range of $0.5 < Q^2 < 1.0$ GeV$^2$.
The transverse part of the in-medium baryon-baryon interaction
entering in the RPA effects
improves the fit, though both are consistent within the errors on the
data, which already include a 10\% reduction of the flux described for
that figure.  Similarly, many of the data points in the angle bins near $\cos\theta = 0$ shown in
Fig. 1 of that paper correspond to a similar region of $Q^2$ but lower muon
energy.  The prediction without RPA effects for those data points is consistently
lower than the data by about 1.5 times the error bar, though if the
10\% reduction in the flux is included both curves would be consistent
with the data.  

\subsection{\label{sec:minerva}MINERvA}

MINERvA's first published results
\cite{Fields:2013zhk,Fiorentini:2013ezn} show a distortion of the
shape of the $Q^2$ spectrum qualitatively similar to other experiments; the
simulation overpredicts the relative rate at low $Q^2$ and underpredicts the
rate at high $Q^2$.  The MINERvA data are presented as an unfolded
differential cross section and a shape relative to the default QE model
from the {\small GENIE} event generator.  The shape comparison has
uncertainties under 10\% because the uncertainty in the flux
substantially cancels out.

\begin{figure*}
\includegraphics[width=8.6cm]{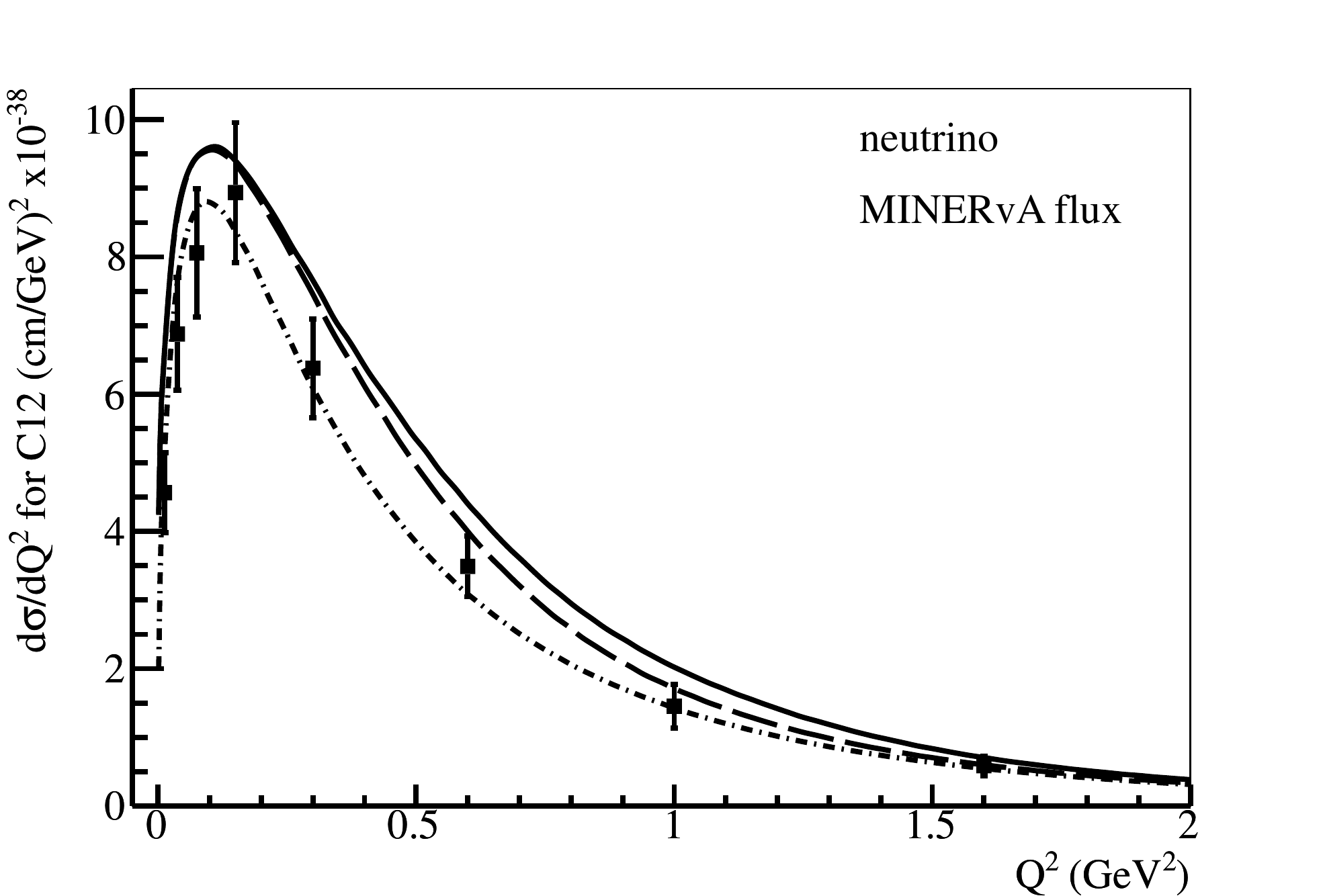}
\includegraphics[width=8.6cm]{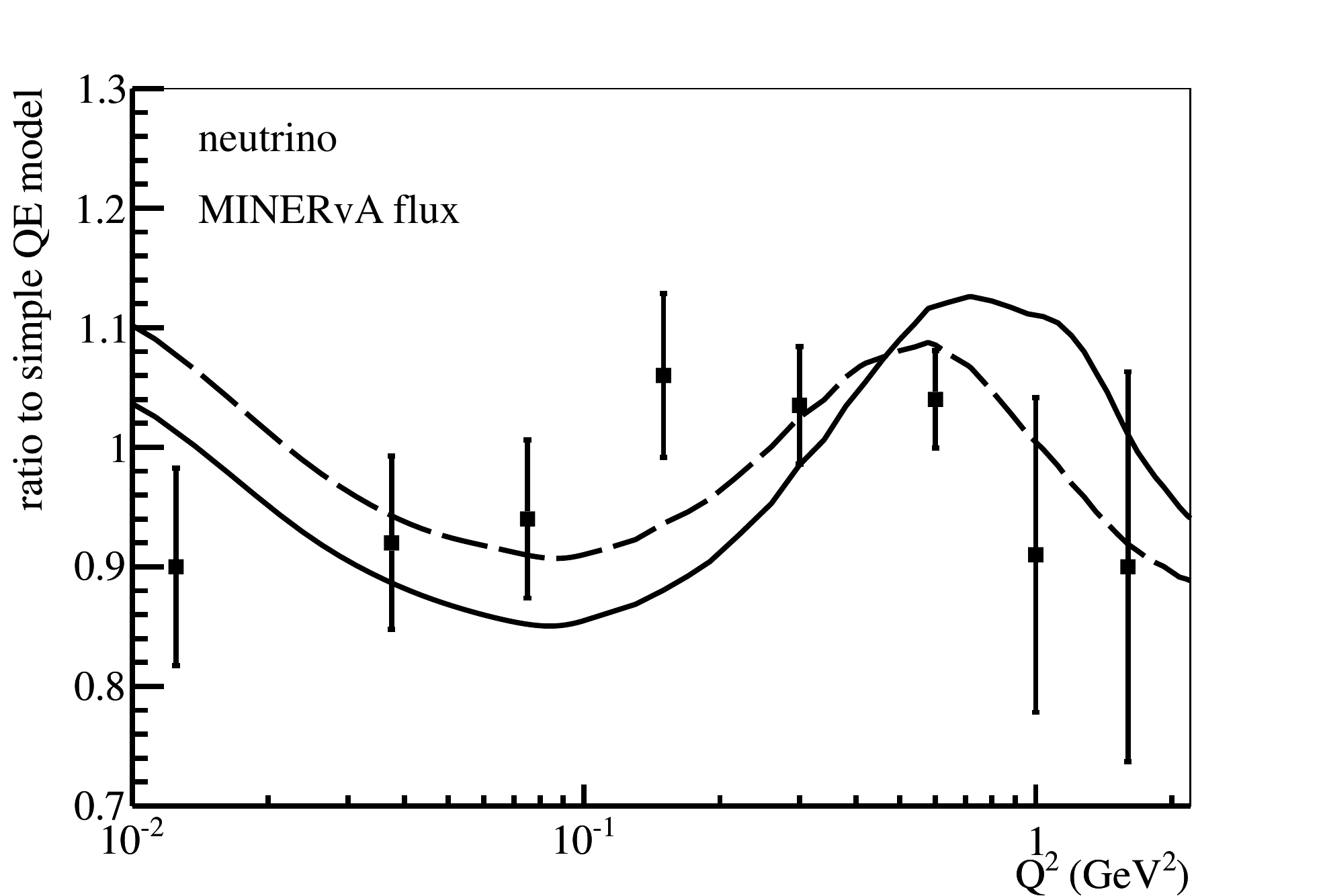}
\includegraphics[width=8.6cm]{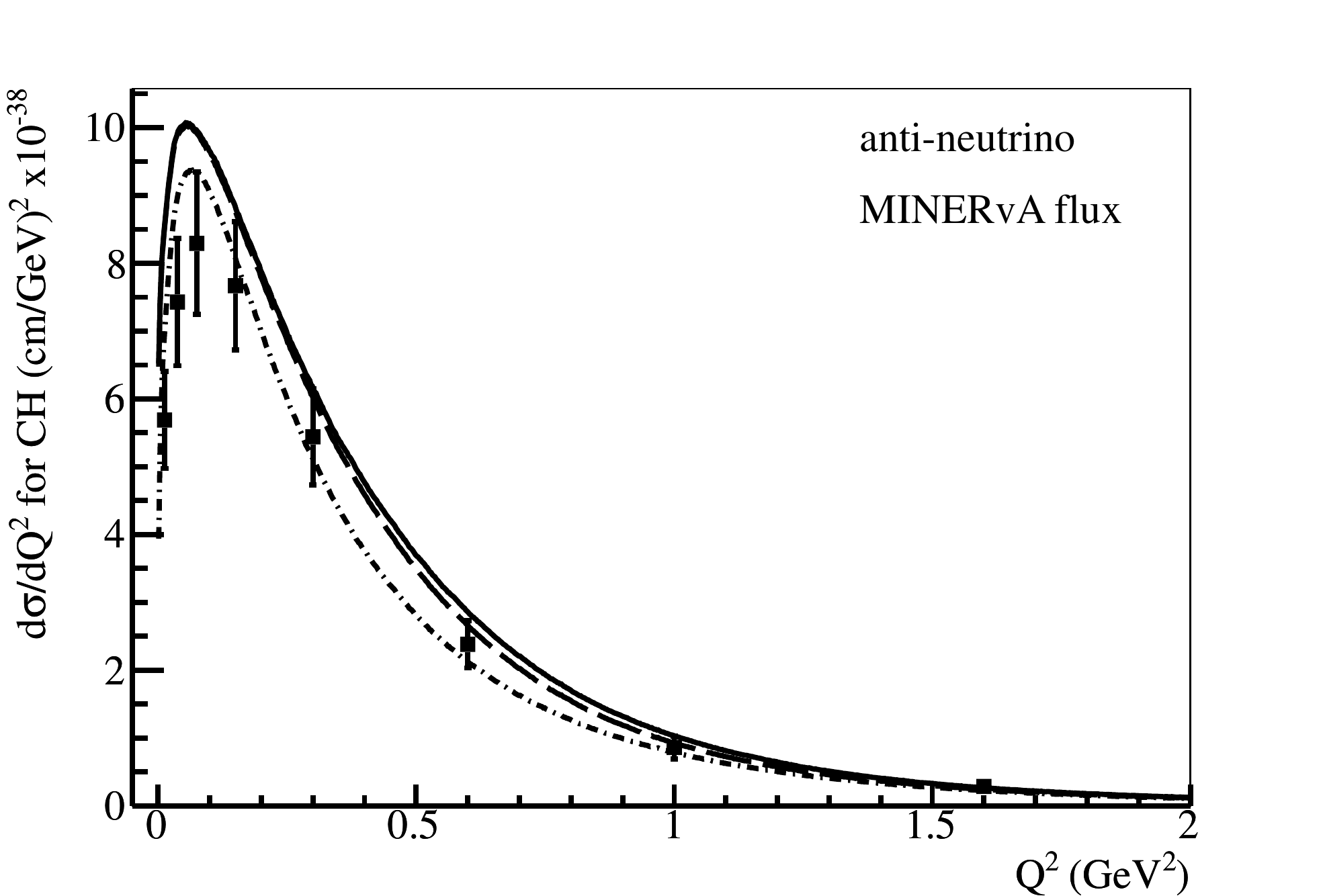}
\includegraphics[width=8.6cm]{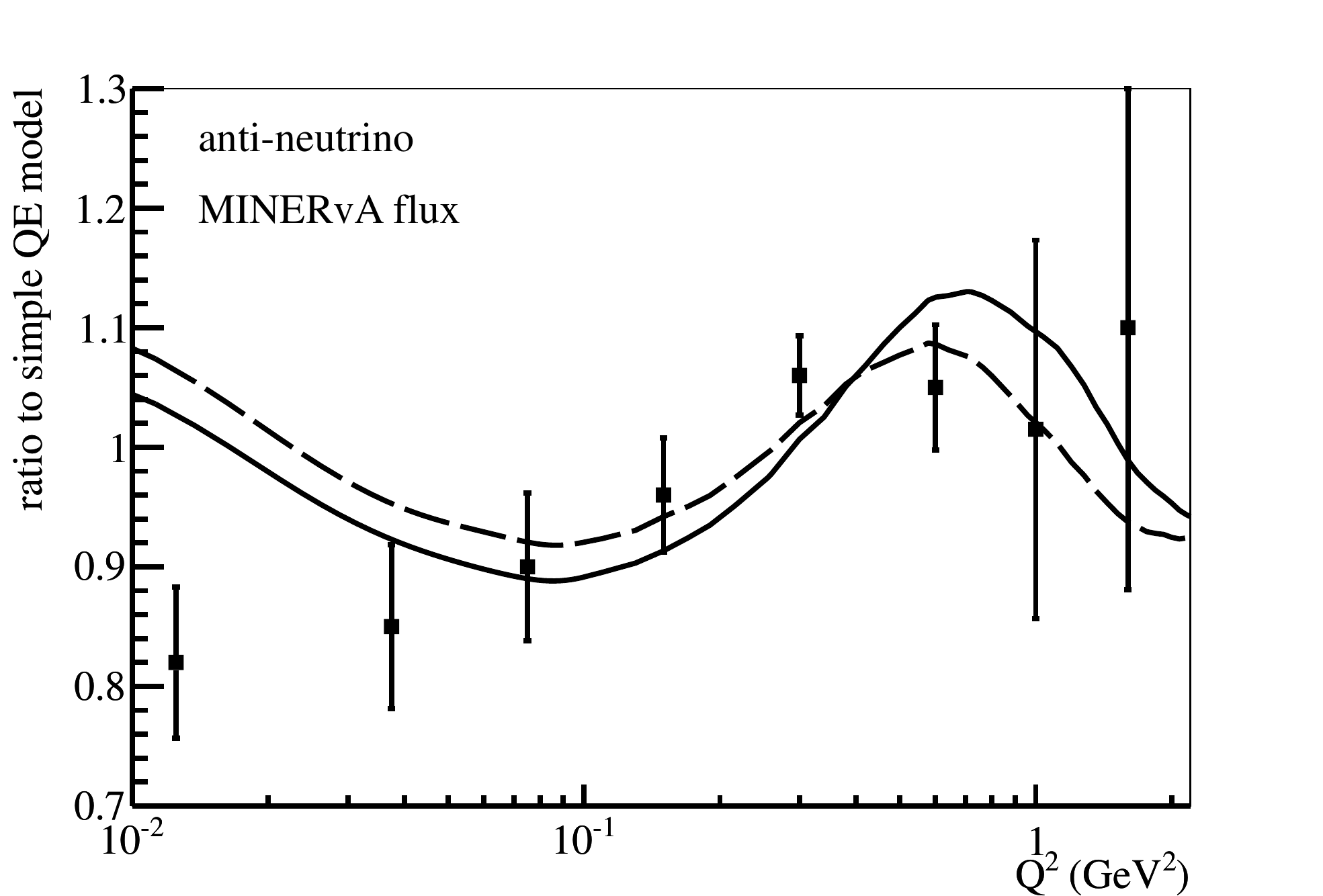}
\caption{\label{fig:minervaqq} Differential $Q^2$ distribution with
  2p2h reconstructed from muon kinematics and QE with RPA effects
  and MINERvA flux
  (solid line) compared to MINERvA data.  Neutrino (top) and antineutrino (bottom)
  with the ratio (right) that reduces several uncertainties especially
  from the flux.  The flux averaged calculation without RPA and without 2p2h is
  shown in the absolute plot (left, dot-dashed line), and is the
  denominator of the calculation in the ratio plot. 
  The ratio for the MINERvA data
  is directly from \cite{Fields:2013zhk,Fiorentini:2013ezn}, which has
  a flux integrated cross section from {\small GENIE} for the denominator
  and the area normalization. The RPA calculation with the alternate high $Q^2$
  dependence is the long-dashed line. 
}
\end{figure*}

Figure~\ref{fig:minervaqq} compares this model to the MINERvA results.
The model is convoluted with the MINERvA flux between 1.5 and 10 GeV.
The appropriate ratio to form for comparison to the MINERvA results in 
Figs. 4 of \cite{Fields:2013zhk,Fiorentini:2013ezn}
uses the flux weighted QE without RPA in the denominator.
The QE with RPA model is shown with the default high $Q^2$ behavior
(solid lines) and again with the alternate behavior (long-dashed
lines) mentioned in Sec.~\ref{Q2}.

For this comparison, the 2p2h $\Delta$ component is included;
the {\small GENIE} model includes pion absorption but not an
additional specific $\Delta N \rightarrow NN$ absorption process, which affects
the size of their background subtraction.
The $Q^2$ distribution for the 2p2h
component is reconstructed from the muon kinematic quantities using
the QE assumption,
just as the MINERvA data and simulated samples are.  The result is an additional
distortion of the $Q^2$ distribution which is still pronounced at 3
GeV, boosting the 2p2h rate at reco $Q^2$ near zero by 20\% and reducing it
by that much at $Q^2 = 1$ GeV$^2$ compared to the true $Q^2$, but the distortion 
is only one-quarter that much at 10 GeV.  There is no significant reco bias
for the QE component implemented in this model, only some additional smearing.

The model describes absolute cross section well.
The area normalized ratio, with reduced flux uncertainties is also
modestly in agreement.
The trend upward with increasing $Q^2$ is
similar, the magnitude of the trend is too large in the default model
but about right for the smaller RPA variation.  Another possible interpretation,
similar to the comparison with the TEM, is that the model peaks at
higher $Q^2$, and more investigation into this behavior might be warranted.

The calculations presented here have not been tuned or modified for higher
energy behavior except for the cut in three-momentum transfer and the
alternate RPA $Q^2$ dependence.  The quality of the MINERvA data and
the uncertainties in the model are such that
5\% to 10\% sized effects are now relevant.   Though detailed
investigation is beyond the scope of this paper, several simple estimates
of effects already mentioned do not individually make the
ratios agree conclusively.  
This includes details specific to the
MINERvA situation:
considering the correlation presented in the MINERvA
systematic uncertainties, energy and angle selection and unfolding
effects, and importantly how much $\Delta$ component
should be included in the comparison.  
On the model side, the QE with no RPA
(dot-dashed line) has a different shape
than {\small GENIE} by $\pm$ 5\%, tuning the QE $M_A$ or form factors
may be reasonable, and a simple estimate of uncertainties related
to the high $Q^2$ behavior of the RPA effects are already presented.

The MINERvA results, especially Figs.~5 in
\cite{Fields:2013zhk,Fiorentini:2013ezn} also 
include the indication that there is an excess
of energy carried by protons in the neutrino case, and little or no such excess
of protons in the antineutrino case.   Though the hadron final state
kinematics are not calculated here, there are two elements that can be
described roughly.  The 2p2h component without the $\Delta$ is expected
to lead to a $pp$ final state half the time, and 5/6 of the time for
the $\Delta$ absorption component, before additional intranuclear rescattering occurs.  For
the antineutrino case, these are the fractions that lead to an $nn$
final state.  Compared to the pure QE process (before rescattering
effects), both pick up additional
protons in the final state if two nucleons leave the nucleus, and give
the lead nucleon or both a little more energy than the QE process.
As noted in the MINERvA papers 
and in \cite{Arrington:2011xs}, the SRC process in electron scattering
is dominated by the $pn$ initial state, which becomes $pp$ final state
for CC neutrino scattering and $nn$ for antineutrino.  Though not
the case for an average over the 2p2h component of the model presented here, the
portion very close to QE kinematics is predicted to similarly be
enriched in the $pn$ initial state.   This
preferentially produces more proton energy for the neutrino case.

\section{\label{sec:conclusion}Conclusion}

A microscopic calculation of the neutrino and antineutrino 2p2h interaction processes
without a pion in the final state produces a cross section that ranges
from 26\% to 29\% of the QE cross
section (14\% to 15\% for the non-$\Delta$ component) 
at energies from 3 and 10 GeV and for isoscalar nuclei
with A $\geq$ 12. For antineutrinos, the range is from 33\% to 32\%
for the full pionless calculation and 18\% to 17\% without $\Delta$ absorption.
These events have a
kinematic signature that is different than QE events, they fill in the
``dip'' region and most would be
reconstructed with systematically low neutrino energy if only lepton
kinematics and the QE assumption is used.  The mix of initial state
for these 2p2h interactions has a complicated dependence, from 50\% to
80\% $pn$ initial state for the non-$\Delta$ and $\Delta$ peaks,
respectively, and a high near QE kinematics. 
The QE cross section is also significantly modified at these energies
especially when RPA calculations of the effect of nuclear correlations
are included.
For an analysis of data describing the shape of the $Q^2$ distribution, this
is likely a larger effect.

This calculation has the 2p2h and RPA effects widely believed to be relevant
and present in electron scattering and also describes the low energy
MiniBooNE data.   Individually, these effects do not modify the simple
QE model in a way that would match the data
but together they qualitatively describe a distortion of the
$Q^2$ spectrum that would likely lead to an anomalous value for the
axial mass parameter for experiments with energies up to 10 GeV.  When
confronted with the MINERvA data and its small uncertainties, 
the model has the qualitative features and magnitude to give
reasonable agreement.   Future MINERvA analyses,
including higher $Q^2$ hadron and 2D muon kinematic distributions,
combined with refinements of the high $Q^2$ part of this model and its application
to the MINERvA situation look very promising.

\smallskip

\begin{acknowledgments}
We wish to thank Boris Popov for discussions and interpretation of
details of the NOMAD analysis, and Panos Stamoulis and Kevin McFarland for discussions of how
these results apply to T2K and MINERvA.  This work has been produced with the
support of the Spanish Ministerio de Econom\'\i a y Competitividad and
European FEDER funds under Contract Nos. FIS2011-28853-C02-01,
FIS2011-28853-C02-02, FPA2011-29823-C02-02, Consolider-Ingenio 2010 Programme CPAN
(CSD2007-0042) and the Severo Ochoa program excellence SEV-2012-0234, 
the Generalitat  Valenciana under Contract No.
PROMETEO/2009/0090 and the E.U. HadronPhysics2 project, Grant
Agreement No. 283286, and the U.S. National Science Foundation
under Grant Nos. 0970111 and 1306944.
\end{acknowledgments}

\bibliography{highe2p2h}

\end{document}